\documentstyle[12pt]{article}

\def\epsffile {}

\if@twoside  m
\else
\fi
\begin{document}

\begin{center}
{\large \bf Band spectra of rectangular graph superlattices}
\vspace{8mm}

P. Exner$^{a,b}$ and R. Gawlista$^c$
\vspace{2mm}

{\small \em $^a$Nuclear Physics Institute, Academy of Sciences, 25068
{\v R}e{\v z} near Prague, Czech Republic} \\
{\small\em $^b$Doppler Institute, Czech Technical University, B{\v
r}ehov{\'a} 7, 11519 Prague, Czech Republic} \\
{\small\em $^c$Lehrstuhl Theoretische Physik I, Fakult{\"a}t f{\"u}r
Physik, Ruhr--Universit{\"a}t Bochum, 44780 Bochum--Querenburg, Germany}
\end{center}
\vspace{3mm}

\begin{quote}
{\small We consider rectangular graph superlattices of sides $\,\ell_1,
\ell_2\,$ with the wavefunction coupling at the junctions either of the
$\,\delta\,$ type, when they are continuous and the sum of their
derivatives is proportional to the common value at the junction with
a coupling constant $\,\alpha\,$, or the $\,\delta'_s\,$ type with the
roles of functions and derivatives reversed; the latter corresponds to
the situations where the junctions are realized by complicated geometric
scatterers. We show that the band spectra have a hidden fractal structure
with respect to the ratio $\,\theta:=\ell_1/\ell_2\,$. If the latter
is an irrational badly approximable by rationals, $\,\delta\,$
lattices have no gaps in the weak--coupling case. We show that there
is a quantization for the asymptotic critical values of $\,\alpha\,$
at which new gap series open, and explain it in terms of
number--theoretic properties of $\,\theta\,$. We also show how the
irregularity is manifested in terms of Fermi--surface dependence on
energy, and possible localization properties under influence of
an external electric field.
}
\end{quote}
\vspace{3mm}

\section{Introduction}

The advent of new technologies which made possible fabrication of
semiconductor quantum wires and other tiny structures opened a new
chapter in solid--state physics as well as ways to many potentially
useful devices. At the same time, this development has an impact
on the quantum theory itself, which is maybe not so spectacular but
by no means less important. The point is that by investigating
various ``taylored" systems one is able to study --- both
theoretically and experimentally --- interesting and sometimes
unsuspected effects ``hidden" in the basic equations of quantum
mechanics.

An example is represented by irregular spectral properties due to
incommensurability of certain parameters of the system. Such behavior
is known to occur, {\em e.g.}, for two--dimensional lattice electrons in
a constant magnetic field.$^{1-5}$
\setcounter{footnote}{5}
In this paper we are going to discuss similar effects which can be
observed {\em without presence of an external field} in certain
graph superlattices.

Before we shall describe the model, a few words should be said about
simplification it involves. A real quantum wire is a complicated
many--body system; even if we suppose that it has an ideal crystallic
structure, we may describe it as an electron duct at most in a certain
range of energies where the profile of the conduction band is
reasonably flat. Neglecting the lateral size of the wire, {\em i.e.},
assuming that the propagating electrons remain in a single transverse
mode, is another approximation. It can be justified in thin wires,
not only by practical experience, but also by rigorous
arguments\footnotemark\ showing that the intermode coupling involves
a dynamical ($\,p$--space) tunneling, and therefore it diminishes
exponentially with the decreasing wire thickness.

Although the replacement of a quantum wire system by the corresponding
graph structure leaves us with a much simpler model, other idealizations
may be useful to draw lessons of it. A typical one concerns the global
size of the system. As the number of individual cells in a superlattice
grows, ``collective effects" become more important; the question is
when they become to prevail. The larger the superlattice, the more
reasonable is to start from an infinitely extended structure, with the
boundary effects considered as a perturbation.

The last introductory remark concerns the question why various results
based on number--theory properties of the parameters represent more
than a nice mathematics. With a certain exaggeration one can certainly
claim that for a mathematician all rationals are the same, while in
physics with its finite--resolution experiments it is meaningless to
ask whether a measured quantity is irrational or not. Fortunately,
differences between number types and between simple and complicated
rationals are usually two sides of the same coin.

While theoretically there might be an ultimate difference between
the value $\,\sqrt 2\,$ and its close rational approximations, it is
manifested over a long scale (in energy, time, {\em etc.}); at a shorter
scale it is important that the value in question  does not coincide
with one of simple harmonies. The irrational model thus represents
a proper description of the ``disharmonic" situation. These somehow
vague statements can be readily illustrated with the help of the system
treated below.

\section{Description of the model}

The configuration space of our model is a two--dimensional lattice
graph whose elementary cell is a rectangle of sides $\,\ell_1,\,
\ell_2\,$ ({\em cf. } Figure~1).
If no external field is applied, the motion of electrons
on graph links is free. Since the choice of the energy scale will
play no role in the following,  we choose atomic units,
$\,\hbar^2/2m^*=1\,$, for the sake of simplicity; hence if the
wavefunction $\,\psi\,$ is supported in the interior of a single
graph link, the Hamiltonian changes it into $\,-\psi^{\prime\prime}\,$.

%
  \begin{figure}
  \begin{center}
  {\phantom{0815}}\vspace{5mm}
  {\hfill\epsffile{Bild1.epsi}\hfill}
  {Figure 1: A rectangular lattice}
  \end{center}
  \end{figure}
%

The nontrivial part of the problem concerns, of course, the behavior
at the junctions. The wavefunctions must be coupled there in such a
way that the probability flow is preserved; in mathematical terms
this is equivalent to the claim that the Hamiltonian is a self--adjoint
operator. This is known for long: Schr\"odinger operators on
graphs appeared in quantum mechanics for the first time
in connection with the free--electron model of organic
molecules,\footnotemark\ and in recent years interest to them has been
revived.$^{8-16}$
\setcounter{footnote}{16}

The requirement of probability--flow  conservation does not specify
the coupling uniquely: it can be satisfied, {\em e.g.}, if the wavefunctions
are continuous at all vertices and their derivatives satisfy there
the conditions
   \begin{equation} \label{delta bc}
\sum_j \psi'_j(x_m)\,=\, \alpha_m\psi(x_m)\,,
   \end{equation}
where $\,m\,$ is the vertex number, the derivatives are taken in the
same direction (conventionally outward), the sum runs over all links
entering this vertex, $\,\psi(x_m)\,$ is the common value of the
functions $\,\psi_j\,$ there, and the real numbers $\,\alpha_m\,$
are coupling constants characterizing the junctions. For the sake of
brevity, we shall refer to ({\rm Re\,}f{delta bc}) as to the
$\,\delta\,$ {\em coupling;} the name is motivated by the fact that
in the simplest case of just two links this is nothing else that the
$\,\delta\,$ interaction on a line.$^2$

A choice of the coupling should, of course, be obtained by deriving
the graph model from a more realistic description, in which the
configuration space consists of a system of coupled tubes. Though
a heuristic argument showing that for an ideal star--like junction
the conditions ({\rm Re\,}f{delta bc}) with $\,\alpha_m=0\,$ might the
optimal choice was given more than four decades ago,$^7$ and it is
natural to expect that nonzero coupling constants could correspond
to local deformation of the junction region, impurities or influence
of external fields --- in short, {\em imperfect contacts} --- no
convicing answer is known up to now.

Moreover, the coupling ({\rm Re\,}f{delta bc}) is not the only possible: for
a junction of $\,N\,$ links there is in general an $\,N^2$ parameter
family of self--adjoint operators which act as the free Hamiltonian
outside the branching points. A method to construct such operators
and some particular classes of them were discussed in detail in Ref.14.
In distinction to ({\rm Re\,}f{delta bc}), the ``additional" couplings have
wavefunctions {\em discontinuous} at the vertex, {\em i.e.}, the limits
for at least one pair of links differ mutually there.

This feature is not automatically disqualifying. It was shown in
Ref.10 that the so--called $\,\delta'$ interaction\footnotemark\
represents a reasonable (if idealized) model for a complicated
geometric scatterer, in which instead by a point contact two
halflines are joined by numerous short ``wires". Moreover, this
result extends to junctions with any number of links;$^{13}$ the
most natural counterpart to ({\rm Re\,}f{delta bc}) appears to be the
so--called $\,\delta'_s\,$ {\em interaction} which requires the
wavefunction {\em derivatives} to be continuous, $\,\psi'_1(x_m)=
\cdots=\psi'_N(x_m)=: \psi'(x_m)\,$, and
   \begin{equation} \label{delta' bc}
\sum_j \psi_j(x_m)\,=\, \beta_m\psi'(x_m)
   \end{equation}
for some $\,\beta_m\,$. The``coupling constants" $\,\beta_m\,$
here measure, roughly speaking, the {\em total length} of the wires
which constitute the geometric scatterer; for a more detailed
discussion we refer to Ref.13.

In what follows we shall be concerned with rectangular lattices
in which the coupling at each junction is the same and belongs
to one the above described types; for the sake of brevity we
shall refer to them as to the $\,\delta\,$ and $\,\delta'_s\,$
lattices, respectively. In a sense, such lattices represent a
generalization of the classical Kronig--Penney model and its
$\,\delta'$ modification$^2$ to higher dimensions.\footnotemark\

\section{General properties of the spectra}

Before proceeding further let us recall some results about the
spectra of the considered lattice Hamiltonians derived in Ref.13.

\subsection{$\,\delta\,$ lattices}

By assumption, a $\,\delta\,$ lattice is a periodic system in both
directions. Performing the Bloch analysis, we arrive at the band
condition\footnotemark\
   \begin{equation} \label{delta band condition}
{\cos\vartheta_1\ell_1-\cos k\ell_1\over \sin k\ell_1}\,+\,
{\cos\vartheta_2\ell_2-\cos k\ell_2\over \sin
k\ell_2}\,-\,{\alpha\over 2k}
\,=\,0 \,.
   \end{equation}
Although an analytic solution can be written in the trivial case
only, the condition ({\rm Re\,}f{delta band condition}) nevertheless allows
to draw many conclusions about the spectrum. Let us rewrite it in
the form
$$
{\alpha\over 2k}\,=\, \sum_{j=1}^2\, {v_j-\cos k\ell_j\over \sin
k\ell_j}\;;
$$
if the quasimomentum components $\,\vartheta_j\,,\; j=1,2\,$, run
though the Brillouin zone, the ranges of the parameters
$\,v_j:=\cos\vartheta_j\ell_j\,$ cover the interval $\,[-1,1]\,$.
It is easy to see that for a fixed $\,k\,$, the maximum of the right
side equals
$$
F_+(k) \,:=\, \sum_{j=1}^2\, \tan\left( {k\ell_j\over
2}\,-\,{\pi\over 2} \left\lbrack {k\ell_j\over\pi} \right\rbrack
\right) \,,
$$
where the the square bracket denotes conventionally the integer part,
and the minimum, $\,F_-(k)\,$, is given by a similar formula with
$\,\tan\,$ replaced by $\,-\cot\,$. It is clear from here that
the gaps of the $\,\delta$--lattice spectrum with a coupling constant
$\,\alpha\,$ on the positive part of the energy axis are determined by
the condition
   \begin{equation} \label{delta gap}
\pm\,{\alpha\over 2k}\,>\, \pm F_{\pm}(k)
   \end{equation}
for $\,\pm\alpha>0\,$, respectively. The negative part of the
spectrum is obtained analogously by comparing
$\,{\alpha\over 2\kappa}\,$ with the extremum values of the function
$\,iF_{\pm}(i\kappa)\,$. Simple consequences of the condition
({\rm Re\,}f{delta gap}) are the following:
   \begin{description}
   \item{(a1)} The spectrum has a band structure; it coincides with
the positive halfline $\,[0,\infty)\,$ if and only if $\,\alpha=0\,$.
   \vspace{-1.8ex}
   \item{(a2)} If $\,\alpha>0\,$, each upper band end is a square
of some $\,k_n:= {\pi n\over\ell_1}\,$ or $\,\tilde k_m:= {\pi
m\over\ell_2}\,$, where $\,n,\,m\,$ are integers. Similarly, for
$\,\alpha<0\,$ each lower band end, starting from the second
one, equals $\,k_n^2\,$ or $\,\tilde k_m^2\,$.
   \vspace{-1.8ex}
   \item{(a3)} The lowest band threshold is positive for
$\,\alpha>0\,$ and negative if $\,\alpha>0\;$; in the case
$\,\alpha< -4(\ell_1^{-1}+\ell_2^{-1})\,$ the whole first band is
negative, and the second one starts at $\,\left(\pi\over L\right)^2\,$,
where $\,L:=\max(\ell_1,\ell_2)\,$.
   \vspace{-1.8ex}
   \item{(a4)} The positive bands shrink with increasing
$\,|\alpha|\,$.
   \vspace{-1.8ex}
   \item{(a5)} Each gap is contained in the intersection of a pair of gaps of
the Kronig--Penney model with the coupling constant $\,\alpha\,$ and
spacings $\,\ell_1\,$ and $\,\ell_2\,$, respectively.
   \vspace{-1.8ex}
   \item{(a6)} All gaps above the threshold are finite. If
there are infinitely many of them, their widths are asymptotically
bounded by $\,2|\alpha|(\ell_1\!+\! \ell_2)^{-1}\!+{\cal O}(r^{-1})\,$,
where $\,r\,$ is the gap number.
   \end{description}
Most of these results have a natural meaning. In particular, (a5) shows
that transport properties of the lattice are better than a combination
of its one--dimensional projections. Notice that the Kronig--Penney
spectral condition$^2$ can be cast into the form ({\rm Re\,}f{delta band
condition}) with a single trigonometric expression on the {\em lhs.}
If an energy value is contained in a Kronig--Penney band in one of the
directions, it is trivially also in a band of the lattice Hamiltonian,
the other factor being annulated by choosing $\,\theta_j=k\,$.
The converse is not true, of course: the condition ({\rm Re\,}f{delta band
condition}) may be satisfied even none of the factors can be
annulated separately. The directions in which the electron is able
to ``dribble" through the lattice will be seen in Section~VI below.

Less trivial is the irregular dependence of the spectrum on the
rectangle--side ratio $\,\theta:=\ell_2/\ell_1\,$ coming from the existence
of competing periods in $\,F_{\pm}(k)\,$. It appears that it is not only
rationality or irrationality of $\,\theta\,$ which matters, but also the
type of irrationality plays a role. In this respect, the situation is
similar to the almost Mathieu equation mentioned in the introduction.

Let us recall some elementary facts from the number theory.\footnotemark\
An irrational number $\,\theta\,$ is {\em badly approximable} by rationals
if there is a positive $\,\delta\,$ such that $\,|q\theta\!-\!p|>\delta q^{-1}$
holds for all integers $\,p,q\,$. There are uncountably many such numbers;
nevertheless, they are rather exceptional in the sense that they form
a zero--measure set. Its complement to the set of all irrationals consists
of numbers which we shall call {\em Last admissible.$^5$} A convenient
way to characterize these number types is through their unique
continued--fraction representations: $\,\theta\,$ is badly approximable
if and only if the infinite sequence of integer coefficients in this
representation is bounded, and Last admissible otherwise. Needless to say,
rationals have finitely many nonzero coefficients.

If $\,\theta\,$ is irrational, the {\em rhs } of ({\rm Re\,}f{delta gap}) is
never zero;
the existence of gaps requires then that there is a subsequence of local
minima which tends sufficiently fast to zero. In this way we were able
in Ref.13 to prove the following results:
   \begin{description}
   \item{(a7)} For a badly approximable $\,\theta\,$ there is $\,\alpha_0>0\,$
such that for $\,|\alpha|<\alpha_0\,$ the spectrum has no gaps above the
threshold.
   \vspace{-1.8ex}
   \item{(a8)} The number of gaps is infinite for any $\,\theta\,$ provided
$\,|\alpha|L> 5^{-1/2}\pi^2\, $; recall that $\,L:=\max(\ell_1,\ell_2)\,$.
   \vspace{-1.8ex}
   \item{(a9)} If $\,\theta\,$ is rational or Last admissible, there are
infinitely many gaps for any $\,\alpha\ne 0\,$.
   \end{description}

The worst irrational in this sense is the golden mean $\,\theta={1\over 2}
(1\!+\!\sqrt 5)\,$ which has the continued--fraction representation
$\,\theta=[1,1,\dots]\,$. In this case the sufficient condition for the
existence of infinitely many gaps is necessary at the same time and
coincides with the critical value of the claim (a7): $\,|\alpha_0|L\,$ is
$\,\pi^2(5\theta)^{-1/2}=3.4699...\;$; more about that will be said
in Section~V below.

\subsection{$\,\delta'_s\,$ lattices}

Replacing the $\,\delta\,$ coupling by the boundary conditions
({\rm Re\,}f{delta' bc}), one can derive the band equation in this
case:$^{13}$
   \begin{equation} \label{delta's band condition}
{\cos\vartheta_1\ell_1+\cos k\ell_1\over \sin k\ell_1}\,+\,
{\cos\vartheta_2\ell_2+\cos k\ell_2\over \sin k\ell_2}\,-\,{\beta
k\over 2} \,=\,0\;;
   \end{equation}
the same argument as above then shows that spectral {\em bands} of
the $\,\delta'_s\,$ lattice with a coupling constant $\,\beta\,$ are
determined by the inequalities
   \begin{equation} \label{delta'_s band}
\mp F_{\mp}(k) \,\ge\,\pm\,{\beta k\over 2}
   \end{equation}
for $\,\pm\beta>0\,$ and $\,k>0\,$, and an analogous relation for
the negative part.

The structure of the spectrum is now different; the condition
(\ref{delta'_s band}) allows to make the following conclusions:
   \begin{description}
   \item{(b1)} The spectrum equals $\,[0,\infty)\,$ if and only if
$\,\beta=0\;$; otherwise there are infinitely many gaps.
   \vspace{-1.8ex}
   \item{(b2)} If $\,\beta>0\,$, the lower end of each band coincides
with some $\,k_n^2\,$ or $\,\tilde k_m^2\,$, where $\,n,\,m\,$ are
integers. The same is true for $\,\beta<0\,$ and the upper band
ends, with the exception of the first one.
   \vspace{-1.8ex}
   \item{(b3)} The lowest band threshold is positive for
$\,\beta>0\,$ and negative if $\,\beta<0\;$; in the case $\,-\ell_1
-\ell_2<\beta<0\,$ the whole first band is negative, and the second
one starts at zero.
   \vspace{-1.8ex}
   \item{(b4)} The positive bands shrink with increasing
$\,|\beta|\,$.
   \vspace{-1.8ex}
   \item{(b5)} Each gap is contained in the intersection of a pair of
$\,\delta'\,$ Kronig--Penney gaps$^2$ with the coupling constant
$\,\beta\,$ and spacings $\,\ell_1\,$ and $\,\ell_2\,$, respectively.
   \end{description}

Instead of the asymptotic behavior (a6) of $\,\delta$--lattices, we have
slightly more complicated result. If a band high in the spectrum is well
separated, its width $\,\Delta_r\,$ is the same as in the $\,\delta'$
Kronig--Penney model, $\,\Delta_r= {8\over \beta\ell_j}+{\cal
O}(r^{-1})\,$. Furthermore, if $\,\theta\,$ is rational and two
bands intersect, $\,k_n=\tilde k_m\,$ for some $\,n,m\,$, we have a
similar expression with $\,\ell_j^{-1}$ replaced by $\,\ell_1^{-1}\!
+\ell_2^{-1}$. It may happen, however, that $\,k_n\,$ and
$\,\tilde k_m\,$ are not identical but close to
each other, so that they still produce a single band. Then the band
width is enhanced; the effect is most profound just before the band
splits. Using the condition (\ref{delta's band condition}), it is
straigtforward to estimate the factor of enhancement by conspiracy
of bands:$^{13}$ asymptotically we have the relation
   \begin{equation} \label{enhancement}
{8\over \beta L}\,+\,{\cal O}(r^{-1}) \,<\, \Delta_r\,<\, {32\over
3\beta}\,(\ell_1^{-1}\!+\! \ell_2^{-1})\,+\,{\cal O}(r^{-1})\,.
   \end{equation}

This brief survey shows, in particular, that despite the generally
irregular pattern, the spectra of $\,\delta\,$ and $\,\delta'_s\,$
lattices inherit the main feature of their Kronig--Peney analogues,
namely that they are dominated asymptotically by bands and gaps,
respectively, at high energies. At the same time, the above listed
results leave many questions open about the actual form of these
spectra and related quantities. In the following sections we are going
to answer some of them.

\section{$\,\delta\,$ lattice spectra}

The form (\ref{delta gap}) of the gap condition makes it possible
to find the $\,\delta\,$ lattice spectra numerically in dependence
on parameters of the model; the results are shown on Figures~2 and 3.
The effect of competing periods is obvious: while a square lattice
has a familiar Kronig--Penney spectrum shape$^2$ with the halved
coupling constant as expected from the gap condition, in the general
case the gap pattern is irregular.

%
  \begin{figure}
  {\phantom{0815}}\vspace{5mm}
  {\hfill\epsffile{Bild2.1.epsi}\hfill}

  {Figure 2: The golden--mean $\,\delta\,$ lattice spectrum
   as a function of the coupling constant. }
  \end{figure}
%
%
  \begin{figure}
  {\phantom{0815}}\vspace{5mm}
  {\hfill\epsffile{Bild3.epsi}\hfill}

  {Figure 3: $\,\delta\,$ lattice spectrum as a function of
   the ratio $\,\theta\,$ for $\,\alpha= 20\,$ (above) and $\,\alpha= -20\,$
   (bottom).}
  \end{figure}
%

The dependence of spectral bands on the ratio $\,\theta\,$ shows
clearly that the gaps are contained in view of (a5) in the
intersections of KP--gaps. The latter can be numbered by an
integer $\,n\,$ such that $\,\pi n\ell_j^{-1}$ is the lower (upper)
gap end for $\,\pm\alpha>0\,$, respectively (with $\,n=0\,$
referring to the region below the bottom of the spectrum$^2$).
This allows us to label the gaps of a $\,\delta\,$ lattice
naturally by a pair $\,(n_1,n_2)\,$ with the integers $\,n_j\,$
tagging the KP--gaps. Notice also that due to (a2) the actual
gaps occupy the lower part of the allowed region if $\,\alpha>0\,$,
and the upper one for $\,\alpha<0\,$. Asymptotically each
intersection is of a parallelepipped form and the actual gap part
represents a diagonal cut of it, which conforms with (a6).

Although the $\,\theta\,$ plots look rather regularly, they contain
a hidden irregular pattern which is revealed if the spectrum is
properly folded. A way to do that is suggested by a relation between
the present model and the theory of multidimensional Jacobi
matrices\footnotemark\ by which $\,\lambda=
2\sin(k(\ell_1\!+\!\ell_2))\,$ is the natural ``folded" energy
variable, the above introduced gap labelling has then a direct
relation to that of Ref.3. The folded spectrum is illustrated on
Figure~4; we plot $\,\lambda\,$
against $\,\ln\theta\,$ to show the symmetry with respect to
the exchange of the basic cell sides. All the gaps close, of course,
and the multiplicity is infinite, but it changes in a fractal
way\footnotemark\ as it is seen when we
plot the regions where the original spectrum has at least one gap.
For the sake of illustrativeness, we delete here the ``regular" $\,(1,1)\,$
gap which produces a concave strip covering a part of the upper half
of the picture. Similar ``non--fractal" contributions come from the
gaps $\,(n_1,1)\,$ and $\,(1,n_2)\,$ with $\,n_j\ge 2\,$ which enter
the picture at sufficiently large $\,|\ln\theta|\,$. Each gap of
the fractal structure ``contains" an infinite series of embedded
gaps as can be seen if we plot just the gap edges; for instance,
the central element $\,(2,2)\,$ of the picture is, in fact,
a sequence $\,(2,2)\supset (3,3)\supset (4,4)\supset\dots\;$.

%
  \begin{figure}
  {\phantom{0815}}\vspace{5mm}
  {\hfill\epsffile{Bild4.1.epsi}\hfill}
  {Figure 4a: The folded $\,\delta\,$ lattice spectrum as a function
   of $\,\ln\theta\,$: The regions with at least one gap;
   the lowest one, (1,1), is deleted.}
  \end{figure}
  \begin{figure}
  {\phantom{0815}}\vspace{5mm}
  {\hfill\epsffile{Bild4.2.epsi}\hfill}
  {Figure 4b: The folded $\,\delta\,$ lattice spectrum as a function
   of $\,\ln\theta\,$: An inset of the previous
   picture.}
  \end{figure}
  \begin{figure}
  {\phantom{0815}}\vspace{5mm}
  {\hfill\epsffile{Bild4.3.epsi}\hfill}
  {Figure 4c: The folded $\,\delta\,$ lattice spectrum as a function
   of $\,\ln\theta\,$: The plot of gap edges in the same scale as (a).}
  \end{figure}
  \begin{figure}
  {\phantom{0815}}\vspace{5mm}
  {\hfill\epsffile{Bild4.4.epsi}\hfill}

  {Figure 4d: The folded $\,\delta\,$ lattice spectrum as a function
   of $\,\ln\theta\,$: Gap edges in a wider scale, with several
   lowest of the
   $\,(n_1,1)\,$ and $\,(1,n_2)\,$ gaps shown.}
  \end{figure}
%

\section{Critical coupling constants}

Due to (a7), the spectrum may contain no gaps if the ratio $\,\theta\,$
is badly approximable and the coupling is weak enough. To get a better
understanding of how the number--theoretic properties of the parameter
imply this result, let us discuss in more detail the ``worst" case of
the golden--mean lattice, $\,\theta={1\over 2} (1\!+\!\sqrt 5)\,$.

Suppose that the coupling constant grows from the zero value. The
condition (\ref{delta gap}) shows that a new gap opens whenever
the hyperbolic graph of the function on the {\em lhs } crosses a local
minimum of the function $\,F_+(\cdot)\,$. These critical values of
the coupling together with the corresponding momenta are shown on
Figure~5. One can make from here several conclusions:

%
  \begin{figure}
  {\hfill\epsffile{Bild5.epsi}\hfill}
  \vspace{5mm}

  {Figure 5: Critical coupling constants for the
           golden--mean $\,\delta\,$ lattice.}
  \end{figure}
%

   \begin{description}
   \item{(i)} Gaps open in series. The asymptotic behavior is
clearly visible and the critical points approach the asymptotical
value from above; hence a new infinite series of gaps opens
``from above" once the coupling constant crosses the next
critical value.
   \item{(ii)} The critical points are extremely sparse; we see
that the patterns are practically equidistant in the logarithmic scale.
This is a nice illustration of the fact that $\,\theta\,$ is badly
approximable: to obtain the next close approximation to it we have to
use integers which are roughly twice as large as their predecessors.
   \item{(iii)} The asymptotic critical values are quantized;  they
appear at multiples of the two basic values $\,{\pi^2\over
\ell\sqrt 5}\, \theta^{\pm 1/2}\,$, where $\,\ell:= \sqrt{\ell_1
\ell_2}\,$. However, not every multiple yields a critical value:
in both series the same sequence of integers repeats, namely
   \begin{equation} \label{quantization}
1,\,4,\,5,\,9,\,11,\,16,\,19,\,20,\,25,\,\dots
   \end{equation}
   \end{description}
Let us attempt to explain this quantization rule. Recall that the
golden mean is approximated by ratios of successive Fibonacci
numbers,$^{17}$ $\;\theta=\lim_{n\to\infty} u_{n+1} u_n^{-1}\,$,
where the $\,u_n\,$ satisfy the recursive relation $\,u_{n+1}=
u_n+u_{n-1}\,$ and assume the values $\,1,2,3,5,8,13,21,\dots\,$
for $\,n=1,2,\dots\,$. In the continued--fraction representation
the $\,n$--th approximant equals $\,\theta_n\!=
[1,\dots\,1,0,\dots]\,$ with $\,n\!+\!1\,$ nonzero coefficients.

Let us ``spoil" the approximation by choosing instead the sequence
of $\,\theta^N_n:= [1,\dots,1,N,0,\dots]\,$ with a fixed positive
integer $\,N\,$ preceded by $\,n\,$ ones. This can be alternatively
written as $\,q_{n+1} q_n^{-1}\,$, where $\,q_n:= Nu_n\!+u_{n-1}\,$,
so that
   \begin{eqnarray*}
q_n^2 \left(\theta-\,{q_{n+1}\over q_n}\right) &\!=\!&
\left\lbrack\, N\, {{\theta^n\!-(-\theta)^{-n}}\over\sqrt 5} \,+\,
{{\theta^{n-1}\!-(-\theta)^{-n+1}}\over\sqrt 5}\, \right\rbrack\,
 (-\theta)^{-n+1}\, \left(1-\, {N\over\theta}\right) \\ \\
&\!\to\!& (-1)^n\, {{1+N-N^2}\over\sqrt 5}
   \end{eqnarray*}
as $\,n\to\infty\,$. The asymptotic threshold values thus are
$$
{\pi^2\over \ell\sqrt 5}\, \theta^{\pm 1/2}\left| N^2\!-N\!-1 \right|
$$
which yields the integer multiples $\,1,5,11,19,29,41,\dots\,$ explaining
a part of the values (\ref{quantization}) but not all of them. The missing
numbers can be obtained if we consider a more general approximation to
the golden mean,
   \begin{equation} \label{gm approximation}
\theta^{A,B}_n\,:=\, {{Au_{n+2}\!+Bu_{n+1}}\over {Au_{n+1}\!+Bu_{n}}}
   \end{equation}
for some positive $\,A,B\,$. The numerical factor in the above limit is
then replaced by $\,|A^2\!-B^2\!-AB|\,$, so the complete sequence
(\ref{quantization}) is obtained when $\,A,B\,$ run through positive
integers. What is important is that $\,A,B\,$ here need {\em not} be
relatively prime. This can be seen directly: it is clear from
(\ref{delta gap}) that replacing approximation sequences
$\,\{q_n\},\, \{p_n\}\,$ by $\,\{Mq_n\},\, \{Mp_n\}\,$ means that the
corresponding local minima of $\,F_+(k)\,$ are enhanced by a factor
which tends to $\,M\,$ as $\,n\to\infty\,$.  These are compared with
$\,{\alpha\over 2k}\,$ at the minima which is {\em divided} by
$\,M\,$, so the asymptotic critical value of the modified sequence is
multiplied by $\,M^2$. In the logarithmic scale the $\,x\,$ coordinates
referring to the modified sequence are just shifted by $\,\log M\;$;
this is also seen on Figure~5.

In this way, we are able to explain how the successive infinite gap
series open in the spectrum as the coupling constant $\,\alpha\,$
increases.

\section{Fermi surfaces}

The above results do not exhaust all possible irregularities of the
model in question. The fact that a given energy value is contained
in a spectral band tells us nothing about the directions in which the
electron may propagate. To this end we have to find what are the
allowed values of the quasimomentum, {\em i.e.}, to determine the Fermi
surface. As mentioned above, the Brillouin zone is the rectangle
   \begin{equation} \label{Brillouin}
{\cal B}\,:=\, \Big\lbrack -{\pi\over\ell_1}\,,\, {\pi\over\ell_1}\Big)\,\times
\,\Big\lbrack -{\pi\over\ell_2}\,,\, {\pi\over\ell_2}\Big)\,,
   \end{equation}
and the Fermi surface is a one--dimensional submanifold in it.
Since it has a four--fold symmetry due to the invarince
of the condition (\ref{delta band condition}) with respect to the
interchanges $\,\vartheta_j \leftrightarrow -\vartheta_j\,$, it is
sufficient to consider a quadrant of $\,{\cal B}\,$. The change of variables
$\,v_j:= \cos\vartheta_j\ell_j\,$ maps it on the square $\,[-1,1]
\times[-1,1]\;$; the band condition becomes at that
$$
{v_1\over\sin k\ell_1}\,+\,{v_2\over\sin k\ell_2}\,=\,
{\alpha\over 2k}\,+\,\cot k\ell_1\,+\,\cot k\ell_2
$$
describing a line segment which crosses the square provided $\,k\,$
corresponds to a point in a spectral band. Passing back to the
quasimomentum coordinates we see that the Fermi surface can have one
of several standard shapes.

On the other hand, the way in which the surface changes as a
function of the Fermi energy depends substantially on the
ratio $\,\theta\,$. This is illustrated on Figure~6 where
we plot the two--dimensional manifolds spanned by the Fermi
surface as the energy increases; to see the change over a larger
range of energies we choose for the latter a logarithmic
scale. We see that for a square lattice the surface ``evolution" produces
a regular pattern of switching ``caps" reminiscent of the corresponding
quantity for the one--dimensional Kronig--Penney model$^2$, while in
the golden--mean case the dependence is highly irregular.

%
  \begin{figure}
  {\phantom{0815}}\vspace{5mm}
  {\hfill\epsffile{Bild6.1.epsi}\hfill}
  {Figure 6a: Fermi surfaces for different energies for the second
band of a square $\,\delta\,$ lattice.}
  \end{figure}
  \begin{figure}
  {\phantom{0815}}\vspace{5mm}
  {\hfill\epsffile{Bild6.3.epsi}\hfill}
  {Figure 6b: Fermi surfaces for different energies, the "side view" for the
same lattice and the logarithmic scale of energy, $\,1\le k\le 21,\;$.}
  \end{figure}
  \begin{figure}
  {\phantom{0815}}\vspace{5mm}
  {\hfill\epsffile{Bild6.2.epsi}\hfill}
  {Figure 6c: Fermi surfaces for different energies for the second
band of a golden--mean $\,\delta\,$ lattice}
  \end{figure}
  \begin{figure}
  {\phantom{0815}}\vspace{5mm}
  {\hfill\epsffile{Bild6.4.epsi}\hfill}

  {Figure 6d: Fermi surfaces for different energies, the "side view" for the
same lattice and the logarithmic scale of energy, $\,1\le k\le 21,\;$.}
  \end{figure}
%

Another lesson from the dependence of Fermi surfaces on energy
concerns the character of the spectrum. To avoid spectral
singularities, the solution $\,k=k(\theta_1,\theta_2)\,$ of the equation
(\ref{delta band condition}) has to be smooth with local extrema at
discrete points only. The {\em lhs } of ({\rm Re\,}f{delta band condition})
which
we denote as $\,D(k,\theta_1,\theta_2)\,$ is certainly smooth in
all variables away of the points $\,k\ell_j=n\pi$, and
$$
{\partial D\over\partial k}(k,\theta_1,\theta_2)\,=\,
{\ell_1\over \sin^2 k\ell_1}\,+\,  {\ell_2\over \sin^2 k\ell_2}\,+\,
{\alpha\over 2k^2} \,\ne\, 0
$$
holds there, with a possible exception of a discrete set of points
for $\,\alpha<0\,$. Hence $\,k(\theta_1,\theta_2)\,$ is smooth by the
implicit--function theorem and
$$
{\partial k\over\partial \theta_j}(\theta_1,\theta_2)\,=\,
-\,\ell_j\, {\sin\theta_j\ell_j \over \sin k\ell_j}\,
\left({\partial D\over\partial k}(k,\theta_1,\theta_2)\right)^{-1}\,,
$$
which may be simultaneously zero only if $\,\theta_j\ell_j=\pm\pi\,,\;
j=1,2\,$. This means that stationary points of $\,k(\theta_1,\theta_2)\,$
are only at the center of the Brillouin zone or at its cornerpoints; we
conclude that
   \begin{description}
   \item{(a10)} the $\,\delta\,$ lattice spectrum in the bands is
absolutely continuous.
   \end{description}
This is not just a mathematical statement; it tells us that the transport
properties of electrons in the allowed energy windows are ``normal".

\section{External fields}

Adding an external field, even a simple one like a linear potential,
makes the problem much more difficult, and we are not able to present
more than several heuristic observations. Recall that the problem
is far to be understood even in the one--dimensional case. It has
been demonstrated recently that an array of $\,\delta'$ interactions
has no absolutely continuous spectrum when an electric field is
applied;$^{10,23}$ this result extends to a wide class of background
potentials.$^{24}$ Moreover, one may conjecture that the spectrum
of $\,\delta'$ Wannier--Stark ladders depends substantially on the
slope of the potential being pure point and nowhere dense if the
latter is rational, and covering the whole real axis otherwise.$^{23}$

The case of a periodic $\,\delta\,$ array is more difficult and only
partial results are known,$^{25-27}$
\setcounter{footnote}{27}
though it is generally believed that a phase transition occurs in
this case with the spectrum being pure point at weak fields and
continuous if the field is strong.$^{26}$ The critical field value
can be conjectured to be $\,F_{\rm crit}\approx \alpha^2/4a\,$, where
$\,a\,$ is the array spacing and $\,F\,$ the field intensity, using the
argument of Ref.26, which is
based on estimating the tunneling probability through the family
of tilted spectral gaps.\footnotemark\ The result is intuitively appealing
because it compares two value having an established meaning: the energy
step $\,Fa\,$ with the bound--state energy $\,\alpha^2/4\,$ of a single
$\,\delta\,$ well. However, the argument neglects the fact that in some
tilted gaps the other (exponentially growing) solution for the
classically forbidden region may play role, and therefore it is not
apriori clear whether this is the correct answer.

Applying the same tilted--band picture to the two--dimensional situation,
one may conjecture that in a $\,\delta'_s\,$ lattice whose spectrum is
dominated by the gaps at high energies, an external electric field
is likely to produce a localization in the field direction; in other
words, electrons would be able to move at most in the direction
{\em perpendicular} to the field.

In the $\,\delta\,$ lattice case the situation is more complicated.
Even if a Berezhkovski--Ovchinnikov--type argument$^{26}$ could be
justified, it will yield now only a lower bound to the power with which
the solution to the Schr\"odinger equation decays, because a gap--width
counting disregards the fact that at some energies belonging to
a band a propagation is possible in certain directions only. With
this reservation in mind, let us compute the expression which is
a two--dimensional analogue of the BO--power: it equals $\,(\pi/
16F)\,{\cal T}(\theta,\alpha)\,$ with
   \begin{equation} \label{T function}
{\cal T}(\theta,\alpha)\,:=\, \lim_{n\to\infty}\,
{1\over \ln n}\:\sum_{m=1}^n\,
{|\delta_m|^2\over k_m}\,,
   \end{equation}
where $\,\delta_m\equiv \delta_m(\theta,\alpha)\,$ are the corresponding
gap widths.

Due to the results of Section~IV, namely the property (a4),
the sum in (\ref{T function}) is monotonously increasing as a function
of $\,|\alpha|\,$, the growth being roughly quadratic. The same
behavior is expected for $\,{\cal T}(\theta,\alpha)\,$ though for a badly
approximable $\,\theta\,$ the existence of the thresholds discussed
in Section~5 may lead to local deformations of this dependence
around the threshold values of $\,\alpha\,$; this is illustrated on
Figure~7.

%
  \begin{figure}
  {\phantom{0815}}\vspace{5mm}
  {\hfill\epsffile{Bild7.1.epsi}\hfill}

  {Figure 7: The quantity $\,{\cal T}(\theta,\alpha)\,$ as a function of
   the coupling constant for $\,\theta= 1,\, {5\over 3},\,
   {233\over 144}\,$, and the golden mean. The arrows mark distinguishable
   gap openings.}
  \end{figure}
%
%
  \begin{figure}
  {\phantom{0815}}\vspace{5mm}
  {\hfill\epsffile{Bild8.3.epsi}\hfill}
  {Figure 8a: The quantity $\,T(\theta,\alpha)\,$ as a function
   of $\,\theta\,$. Plot for $\,\alpha=3.4699\,$ over rational
   values of $\,\theta\,$.}
  \end{figure}
  \begin{figure}
  {\phantom{0815}}\vspace{5mm}
  {\hfill\epsffile{Bild8.4.epsi}\hfill}
  {Figure 8b: The quantity $\,T(\theta,\alpha)\,$ as a function
   of $\,\theta\,$. Plot for $\,\alpha=3.4699\,$ with $\,\theta\,$ being random
identically
   distributed numbers.}
  \end{figure}
  \begin{figure}
  {\phantom{0815}}\vspace{5mm}
  {\hfill\epsffile{Bild8.1.epsi}\hfill}
  {Figure 8c: The quantity $\,T(\theta,\alpha)\,$ as a function
   of $\,\theta\,$. Plot for $\,\alpha=20\,$ over rational
   values of $\,\theta\,$.}
  \end{figure}
  \begin{figure}
  {\phantom{0815}}\vspace{5mm}
  {\hfill\epsffile{Bild8.2.epsi}\hfill}

  {Figure 8d: The quantity $\,T(\theta,\alpha)\,$ as a function
   of $\,\theta\,$. Plot for $\,\alpha=20\,$ with $\,\theta\,$ being random
identically
   distributed numbers.}
  \end{figure}
%

On the other hand, let us fix the coupling constant at two
values, $\,\alpha=3.4699...\,$ for which we know that
$\,{\cal T}(\theta,\alpha)\,$ has a zero, and $\,\alpha=20\,$, and plot
the $\,\theta\,$ dependence (Figure~8). Simple harmonic values
of the ratio produce well pronounced peaks; hence the corresponding
lattices are expected to exhibit better localization properties when
an electric field is applied. On the other hand,
$\,{\cal T}(p/q,\alpha)\,$ may depend substantially on the integers
$\,p,q\,$ involved. To illustrate the difference we compute first
this quantity for $\,p,q=1,2,3,\dots\,$ with $\,q\le 150\,$, and
after that we plot $\,{\cal T}(\theta,\alpha)\,$ with random identically
distributed $\,\theta\,$ values; the latter are almost surely
``typical", {\em i.e.}, irrational.

\section{$\,\delta'_s\,$ lattice spectra}

Let us finally check how the spectra of $\,\delta'_s\,$ lattices
look like. They can be found with the help of the condition
(\ref{delta'_s band}); in the same way as above, we can also
check that
   \begin{description}
   \item{(b6)} the $\,\delta'_s\,$ lattice spectrum in the bands is
absolutely continuous.
   \end{description}
Figure~9 shows the dependence of spectral
bands on $\,\beta\,$, again for the golden--mean case,
$\,\theta={1\over 2}(1\!+\!\sqrt 2)\,$; we see that the pattern
has {\em globally} the same behavior as the $\,\delta'$
Kronig--Penney spectrum.$^2$ On the other hand the $\,\theta\,$ plot
(Figure~10) illustrates the properties (b2) and (b5); the latter can be again
used to label the gaps. Moreover, we see clearly the enhancement
due to conspiracy of bands -- {\em cf. } (\ref{enhancement}).

%
  \begin{figure}
  {\phantom{0815}}\vspace{5mm}
  {\hfill\epsffile{Bild9.2.epsi}\hfill}

  {Figure 9: The coupling constant plot for the $\,\delta'_s\,$
   lattice spectrum in the golden mean case.}
  \end{figure}
%
%
  \begin{figure}
  {\phantom{0815}}\vspace{5mm}
  {\hfill\epsffile{Bild10.2.epsi}\hfill}

  {Figure 10: The $\,\delta'_s\,$ lattice spectrum as a function of
   the $\,\theta\,$ for $\,\beta= 5\,$.}
  \end{figure}
%
%
  \begin{figure}
  {\phantom{0815}}\vspace{5mm}
  {\hfill\epsffile{Bild11.epsi}\hfill}

  {Figure 11: The folded $\,\delta'_s\,$ lattice spectrum as a function of
   the $\,\theta\,$ for $\,\alpha= 20\,$. The value of momentum is restricted
   to $\,k\le 100\,$.}
  \end{figure}
%
%
  \begin{figure}
  {\phantom{0815}}\vspace{5mm}
  {\hfill\epsffile{Bild12.1.epsi}\hfill}
  {Figure 12a: Folded--spectrum {\em gaps} for rational approximations to
   the golden mean: $\,\beta=20,\,$.}
  \end{figure}
  \begin{figure}
  {\phantom{0815}}\vspace{5mm}
  {\hfill\epsffile{Bild12.2.epsi}\hfill}

  {Figure 12b: Folded--spectrum {\em gaps} for rational approximations to
   the golden mean: $\,\beta=5\,$.}
  \end{figure}
%

As in the $\,\delta\,$ situation, the spectrum may be
folded into $\,2\sin(k(\ell_1\!+\!\ell_2))\,$; this is shown on
Figure~11. We have deleted here again the lowest ``regular" band
which can be done unambigously if $\,|\ln\theta|\,$ is small enough.
In this case the folded spectrum leaves, in general,
some gaps open, but it may happen at rational values
of $\,\theta\,$ only. Indeed, it is not difficult to check
that for an irrational $\,\theta\,$ spectrum the folded spectrum has
a full measure. It is sufficient to notice that all the points
$\,{\pi n\over\ell}\,\theta^{\pm 1/2}$ with both signs are
contained in a band, either as its lower edge (upper edge for
$\,\beta<0\,$) or an internal point. The corresponding values
of $\,\lambda\,$ are then $\,2 \sin(\pi n(1\!+\!\theta^{\pm 1}))\,$;
they cover the interval $\,[-2,2]\,$ due to the basic result of
the ergodic theory.\footnotemark\ On the other hand, for rational
$\,\theta\,$ some gaps remain open; this is illustrated on Figure~12.

\section{Conclusions}

We have demonstrated irregular properties of rectangular
$\,\delta\,$ and $\,\delta'_s\,$ lattices with incommensurate sides,
as they are manifested in the form of their spectra, critical
coupling--constant values, dependence of the Fermi surface on
the Fermi energy, and possibly in localization properties under
influence of external fields. These features are expected
to play a dominating role in behavior of finite but large
enough graph superlattices. It is also natural to conjectures
that similar geometrically induced spectral properties will
be seen in superlattices in which the basic cell has another form.

One lesson drawn from the present considerations concerns the
validity of the so--called {\em Bethe--Sommerfeld conjecture} by which
the number of gaps in periodic systems of dimension two or higher
is finite. This was proven by Skriganov\footnotemark\ for
Schr\"odinger operators with nice potentials, and it is known
to be valid also for some systems with $\,\delta\,$ interactions.$^2$
We have seen that the conjecture may not be valid if the
periodic structure is singular enough: the golden--mean
$\,\delta\,$ lattice has, depending on the value of the
coupling constant $\,\alpha\,$, either infinitely many gaps or
{\em none at all.}

\vspace{5mm}

{\noindent}
The work was done during the visits of P.E. at the Institut f{\"u}r
Mathematik, Ruhr--Universit{\"a}t Bochum and R.G. at the Nuclear
Physics Institute, Czech Academy of Sciences; the authors express
their gratitude to the hosts.
The research has been partially supported by the the Grants AS
No.148409 and GACR No.202--93--1314, and the European Union Project
ERB--CIPA--3510--CT--920704/704.

   \footnotetext[1]{The literature about this subject is huge; let
us just mention the classical papers:
P.G.~Harper, Proc.Roy.Soc.(London) A {\bf 68}, 874 (1955);
D.R.~Hofstadter, Phys.Rev. B {\bf 14}, 2239 (1976).
For a more complete bibilography to this problem, which is usually
referred to as the ``almost Mathieu equation", see Ref.2, Sec.III.2.5
or Ref.3; recent mathematical results are discussed, {\em e.g.}, in
Refs.4 and 5.
}
   \footnotetext[2]{S. Albeverio, F. Gesztesy, R. H{\o}egh--Krohn, H.
Holden: {\em Solvable Models in Quantum Mechanics,} Springer--Verlag,
Heidelberg 1988.}
   \footnotetext[3]{J. Bellissard,
in {\em Number Theory and Physics,} (M. ~Waldschmidt et al., eds.),
Springer, Heidelber 1992; pp. 538--630.
}
   \footnotetext[4]{M.A. Shubin, Commun.Math.Phys. {\bf 164}, 259 (1994).
}
   \footnotetext[5]{Y.~Last, Commun.Math.Phys. {\bf 164}, 421 (1994).
}
   \footnotetext[6]{P.~Duclos, P.~Exner, P.~\v S\v tov\'\i\v cek,
Ann.Inst.H.Poincar\'e: Phys.Th\'eor. {\bf 62}, 81 (1995)
}
   \footnotetext[7]{K.~Ruedenberg, C.W.~Scherr, J.Chem.Phys. {\bf 21},
1565 (1953).
}
   \footnotetext[8]{V.M.~Adamyan, Oper.Theory: Adv. Appl. {\bf 59}, 1 (1992).
}
   \footnotetext[9]{Y.~Avishai, J.M.~Luck, Phys.Rev. B {\bf 45}, 1074 (1992).
}
   \footnotetext[10]{J.E.~Avron, P.~Exner, Y.~Last, Phys.Rev.Lett.
{\bf 72}, 896 (1994).
}
   \footnotetext[11]{
J.E.~Avron, A.~Raveh, B.~Zur, Rev.Mod.Phys. {\bf 60}, 873 (1988).
}
   \footnotetext[12]{W.~Bulla, T.~Trenckler, J.Math.Phys. {\bf 31},
1157 (1990).
}
   \footnotetext[13]{P.~Exner, Phys.Rev.Lett. {\bf 74}, 3503 (1995);
J.Phys. A, {\em submitted for publication}
}
   \footnotetext[14]{P.~Exner, P.~\v{S}eba, Rep.Math.Phys. {\bf 28},
7 (1989).
}
   \footnotetext[15]{J.~Gratus, C.J.~Lambert, S.J.~Robinson, R.W.~Tucker,
J.Phys. A {\bf 27}, 6881 (1994).
}
   \footnotetext[16]{N.I.~Gerasimenko, B.S.~Pavlov, Sov.J.Theor.Math.Phys.
{\bf 74}, 345 (1988).
}
   \footnotetext[17]{A single--center $\,\delta'$ interaction is defined
by the requirement of the wavefunction--derivative continuity together
with the condition $\,\psi(0+)-\psi(0-)= \beta\psi'(0)\,$ for some ``coupling
constant" $\,\beta\,$. A thorough discussion of its properties can be found
in Ref.2.}
   \footnotetext[18]{Considerations of the present paper allow a
straightforward extension to graph lattices in dimension $\,d\ge 3\,$.}
   \footnotetext[19]{The condition (\ref{delta band condition}) was obtained
in Ref.6 for $\,\alpha=0\,$ and extended to the nonzero--coupling case in
Ref.12. These authors pursued other goals, however, and therefore restricted
themselves to writing down the solution in the trivial situation.}
   \footnotetext[20]{G.H.~Hardy, E.M.~Wright, {\em An Introduction to the
Theory of Numbers,} Oxford University Press 1979; W.M.~Schmidt, {\em
Diophantine Approximations and Diophantine Equations,} LNM 1467,
Springer--Verlag, Berlin 1991.
}
\footnotetext[21]{P. Exner, {\em submitted for publication}}
\footnotetext[22]{We use this term in a somewhat loose sense without
insisting that the folded--gap pattern is exactly self--similar.}
\footnotetext[23]{M.~Maioli, A.~Sacchetti, J.Phys. A {\bf 28}, 1101 (1995)
}
\footnotetext[24]{P.~Exner, J.Math.Phys. {\bf 36}, {\em to appear}
}
\footnotetext[25]{F.~Bentosela, V.~Grecchi, Commun.Math.Phys. {\bf 142},
169 (1991),
and references therein.
}
\footnotetext[26]{A.M.~Berezhkovski, A.A.~Ovchinnikov, Sov.Phys.: Solid State
{\bf 18}, 1908 (1976)
}
\footnotetext[27]{P.~Ao, Phys.Rev. B {\bf 41}, 3998 (1990)
}
\footnotetext[28]{To get the stated result from this argument, a
numerical factor decribing the tunneling through a single gap must
be properly evaluated as pointed out by F.~Bentosela.}
   \footnotetext[29]{I.P.~Cornfeld, S.V.~Fomin, Ya.G.~Sinai:
{\em Ergodic Theory}, Springer, New York 1982; Thm.7.2.1}
\footnotetext[30]{M.M.~Skriganov, Sov.Math.Doklady {\bf 20}, 956 (1979);
M.M.~Skriganov, Invent.Math. {\bf 80}, 107 (1985);
see also Yu.E.~Karpeshina,
Proc. Steklov Math.Inst., Issue 3, 109 (1991).
}

\end{document}